\documentstyle[pra,aps,preprint]{revtex}

\begin{document}
\draft

\title{Counter-example where cosmic time keeps its original role in
quantum cosmology\footnote{The talk presented to International
colloquium on the science of time "Time and Matter" at Venice
International University, Italy, August 11-17, 2002}}

\author{E. I. Guendelman \thanks{guendel@bgumail.bgu.ac.il} and
A.  B.  Kaganovich \thanks{alexk@bgumail.bgu.ac.il}}
\address{Physics Department, Ben Gurion University of the Negev, Beer
Sheva
84105, Israel}
\maketitle
\begin{abstract}

In the minisuperspace models of quantum cosmology, the absence of
time in the Wheeler-DeWitt (constraint) equation,
 is
the main point leading to the generally accepted conclusion
that in the quantum cosmology there is no possibility to describe
 the evolution of the universe procceding in the cosmic time
(the time usually used in classical cosmology). 
We show that in spite of the
constraint, under the specific circumstances, the averaging of some
of the Heisenberg
equations can give
nontrivial additional information about {\it explicit time dependence of
the expectation values of certain dynamical variables in  quantum
cosmology}. This idea is realized 
explicitly in a higher dimensional model with a negative 
cosmological constant and dust as the sources of gravity. 
When there is an anisotropy in the evolution of the universe,
 the above phenomenon (i.e. explicit cosmic time dependence of
certain expectation values) appears and  we find the new quantum
effect which consists in 
"quantum inflationary phase" for some dimensions and simultaneous 
"quantum deflationary contraction" for the remaining
dimensions. The expectation value of the "volume" of the universe remains 
constant during this quantum "inflation-deflation"process.

 \end{abstract}

    \renewcommand{\baselinestretch}{1.6}

\pagebreak

\section{Introduction}

In the quantization of generally covariant systems, like General
Relativity (in any number of dimensions), one has to take into account a
fundamental constraint which basically tell us that the total Hamiltonian
of the classical system equals zero\cite{Kuchar}, $H\approx 0$. The sign
$\approx$
is used\cite{Dirac} instead of $H=0$ to emphasize the fact that although
$H$ is zero along the classical trajectories, it is still non trivial in a
sense that it may have non zero Poisson brackets with other dynamical
variables. 

In the quantized version  of the theory, the physical states have to
satisfy the constraint equation $H\Psi =0$ which means that those states
should be time independent. 
One believes usually that
the expectation values
of the Heisenberg equations for operator of {\it any} dynamical variable
is always
zero just by virtue of the constraint equation, i.e. expectation values of
all the dynamical
variables must be time independent too.
Such a situation is interpreted  
in the literature as the statement (having almost the power of a theorem)
that time disappears from quantum
gravity\cite{Kuchar,Sm,Hall}. In particular, in the
context 
of quantum cosmology, this statement is formulated as the  generally
accepted conclusion
that there is no possibility to describe
 the evolution of the universe procceding in the cosmic time.

We will see here that the above conclusions can be premature. This will be
done
by presentation of an explicit counter-example where cosmic time keeps
its original role in
quantum cosmology: the expectation values of certain variables have non
trivial dependence on the same cosmic time which enters, for instance, in
the
classical equations of the Friedmann cosmology. 

The reason for the recovery of a non trivial time dependence in quantum
cosmology, in spite of the fact that $H$ equals zero when applied on
physical states, lies on the two complementary facts: 
a) The Hamiltonian is still a nontrivial operator since it must have non
trivial commutators in order to reproduce the Heisenberg equations. This
means that there must be non physical states $|N.P.\rangle $ for which
$H|N.P.\rangle\neq 0$. \quad b) In our specific model, the physical
states $\Psi$, for which    
$H|\Psi\rangle = 0$ is satisfied, are found to be non normalizable states.
This forces us to consider in any case the "non physical" states of a)
in order to define expectation values of relevant operators with the help
of a limit process $|N.P.\rangle\to|\Psi\rangle$. Since for the non
physical states $H|N.P.\rangle\neq 0 $,  non trivial time dependence can
appear in the the expectation
values
of some of the Heisenberg equations.   

We will display this interesting phenomenon of the appearance of
nontrivial time dependence in quantum cosmology in the context of a
Kaluza-Klein model which allows for anisotropic evolution: expansion of
3 dimensions and contraction of the extra dimensions for a physically
attractive scenario. In this model, a negative cosmological constant does
not let the total volume of the universe grow, while quantum effects
stabilize the volume against collapse.

It is important to point out that the appearance of time we display in our
model,
{\it does not} rely  on some WKB approximation\cite{} or on the use of
some field variable as time\cite{}, rather, it is the {\it genuine,
original cosmic time} which fulfils its natural duty to be the time
parameter of the theory even at the intrinsic quantum level.  
A somewhat related approach which also gives  cosmic time dependence
for averages of certain dynamical variables  in the presence of anisotropy
was developed by Kheyfets and Miller in Ref.$[5]$. For the
generalization of this approach see Ref.$[6]$.

\section{Description of the model and its classical dynamics}
We start from studying a higher dimensional homogeneous totally
anisotropic spatially flat cosmological model
\begin{equation}
ds^2 =-dt^2 +\sum_{l=1}^{D}a_{l}^{2}(t)dx_{l}^{2}
\label{metric}
\end{equation}
which is assumed to be toroidally compact: $0\leq x_{l}\leq L_{l}$. The
scalar curvature corresponding to Eq.(\ref{metric}) is given by
\begin{equation}
R=\frac{2}{V}\frac{d^{2}V}{dt^{2}}-\frac{D-1}{D}\left(\frac{d(\ln
V)}{dt}\right)^{2}+\frac{1}{D}\sum_{l<m}\left(\frac{d(\ln
a_{l})}{dt}-\frac{d(\ln a_{m})}{dt}\right)^{2}
\label{R}
\end{equation}
where $V=\prod_{l=1}^{D}a_{l}\equiv $ "volume" of the universe.
We will assume that the only sources of gravity are a negative cosmological
constant $\Lambda <0$ and dust. The gravitational action can be written
then as
\begin{equation}
S_{gr}=\frac{1}{\kappa}\int d^{D}xdt\sqrt{-g}(R-2\Lambda)
=-\frac{1}{\kappa}\prod_{l=1}^{D}L_{l}\int Ldt
\label{Sgrav}
\end{equation}
where $\kappa = 16\pi G$ and up to a total derivative term, the Lagrangian
$L$ is given by
\begin{equation}
L=\frac{1}{V}\frac{D-1}{D}\left(\frac{dV}{dt}\right)^{2}-
\frac{V}{D}\sum_{l<m}\left(\frac{d(\ln
a_{l})}{dt}-\frac{d\ln (a_{m})}{dt}\right)^{2}
+2\Lambda V
\label{L}
\end{equation}

For simplicity, we choose units where
$\frac{1}{\kappa}\prod_{i=1}^{D}L_{l}=1$ so that
\begin{equation}
S_{gr}=-\int Ldt
\label{Sgrav1}
\end{equation}

In addition to the equations derived from (\ref{L}) and (\ref{Sgrav1}) we
have to impose the constraint saying that the Hamiltonian $H$ is zero (the
statement which coincides with the $0-0$ component of Einstein's
equations). The presence of dust affects only the constraint equation
which
reads
\begin{equation}
H=\frac{1}{V}\frac{D-1}{D}\left(\frac{dV}{dt}\right)^{2}-
\frac{V}{D}\sum_{l<m}\left(\frac{d(\ln
a_{l})}{dt}-\frac{d(\ln a_{m})}{dt}\right)^{2}
-2\Lambda V -\mu =0,
\label{H}
\end{equation}
where $\mu >0$ is a constant which has the interpretation of the dust
energy density times the volume V of the universe.

The form of the Lagrangian and the constraint may be simplified to a
marked degree if one uses the following parametrization for $a_{l}(t)$
\begin{equation}
a_{l}(t)=[V(t)]^{1/D}e^{\theta_{l}(t)}.
\label{teta-param}
\end{equation}
Since $\sum_{l=1}^{D}\theta_{l}\equiv 0$ one can exclude $\theta_{D}
=-\sum_{i=1}^{D-1}\theta_{i}$ and proceed with the $D$ independent 
variables: $ V $
and
$\theta_{i}$, \, $i=1,2,...D-1$.
Finally, one can see that the Lagrangian and
the constraint take the diagonalized and normalized form
\begin{equation}
L=\left(\frac{d\rho}{dt}\right)^{2}-
\rho^{2}\sum_{i=1}^{D-1}\left(\frac{dz^{i}}{dt}\right)^{2}-
\omega^{2}\rho^{2},
\label{L-diag}
\end{equation}
\begin{equation}
H=\left(\frac{d\rho}{dt}\right)^{2}-
\rho^{2}\sum_{i=1}^{D-1}\left(\frac{dz^{i}}{dt}\right)^{2}+
\omega^{2}\rho^{2}-\mu =0,
\label{H-diag}
\end{equation}
where
\begin{equation}
\omega^{2}=-\frac{D}{2(D-1)}\Lambda ; \quad   \rho^{2}=\frac{4(D-1)}{D}V
\label{rho}
\end{equation}
and
\begin{eqnarray}
z^{i}=\frac{1}{1+\sqrt{D}}\sqrt{\frac{D}{2(D-1)}}[\theta_{1}+\theta_{2}+...
+(2+\sqrt{D})\theta_{i} +...+\theta_{D-1}], 
\nonumber\\
 i=1,2,...,D-1
\label{z}
\end{eqnarray}

Notice that the Lagrangian (\ref{L-diag}) and the constraint
(\ref{H-diag}) are invariant under $\frac{D(D-1)}{2}$ dimensional symmetry   
group of
translations and rotations of a $D-1$ dimensional Euclidean space.
In particular, the translational symmetry $z^{i}\rightarrow z^{i}+b^{i}$
with constants $b^{i}$, \, $i=1,2,...,D-1$, gives rise to $D-1$ conserved
quantities
\begin{equation}
F_{i}=-2\rho^{2}\frac{dz^{i}}{dt}, \quad i=1,2,...,D-1
\label{Fi}   
\end{equation}

As we will see in what follows, it is very important that the set of
quantities $\frac{dz^{i}}{dt}$ (which are the linear combinations of
$\frac{d\theta ^{i}}{dt}$) measure anisotropy of the evolution of the
universe. In fact, since  $\sum_{i=1}^{D}\theta_{i}\equiv 0$, all
$\theta_{i}$'s can have the same time dependence only if such time
dependence is the trivial one, i.e., all of the $\theta_{i}$'s are
constants. Therefore, the set of $D-1$ conserved independent quantities
$F_{i}$  measure anisotropy of
the evolution of the universe.

It is interesting to note that the constraint
may be represented now in the following
form
\begin{equation}
H=\left(\frac{d\rho}{dt}\right)^{2}+U^{(class)}_{eff}(\rho)-\mu =0
\label{constr-potential}  
\end{equation}
where the classical effective volume dependent potential appears:
\begin{equation}
U^{(class)}_{eff}(\rho)\equiv -\frac{F^{2}}{4\rho^{2}}+\omega^{2}\rho^{2},
\quad F^{2}\equiv\sum_{i=1}^{D-1}F_{i}^{2}
\label{Ueff}
\end{equation}
All classical solutions exhibit cosmological singularities. This feature,
as we will see below, can be avoided in quantum cosmology in the presence
of
negative
cosmological constant and dust.

\section{Minisuperspace quantization and solutions of the Wheeler-DeWitt
equation}

To produce a quantum theory from the classical one, we have to
postulate canonical commutation relations and in addition, the constraint    
equation (\ref{H-diag}) has to be imposed as a condition on the wave
function of the universe $\Psi$. The resulting Wheeler-DeWitt equation
$H\Psi =0$ is the
fundamental equation governing the quantum cosmology.

It is convenient to rewrite the Lagrangian (\ref{L-diag}) and the
Hamiltonian (\ref{H-diag}) in a geometrical fashion   
\begin{equation}
L=f_{\alpha\beta}(q)\frac{dq^{\alpha}}{dt}\frac{dq^{\beta}}{dt}-
\omega^{2}\rho^{2},
\label{Lgeom}
\end{equation}
\begin{equation}
H=f_{\alpha\beta}(q)\frac{dq^{\alpha}}{dt}\frac{dq^{\beta}}{dt}+
\omega^{2}\rho^{2}-\mu =\frac{1}{4}f^{\alpha\beta}\pi_{\alpha}\pi_{\beta}+
\omega^{2}\rho^{2}-\mu
\label{Lgeom}
\end{equation}
where
\begin{equation}
q^{\alpha}=(\rho ,z^{i} ), \quad i=1,2,...,D-1; \quad
f_{\alpha\beta}(q)= diag(1, -\rho^{2},...,-\rho^{2})
\label{metric}
\end{equation}
are respectively
 coordinates and metric of our minisuperspace and $\pi_{\alpha}$ are the
momenta canonically conjugate to $q^{\alpha}$. 

The
ambiguity due to
operator ordering problem can be taken into
account\cite{Kuchar1}
by adding the "nonminimal" term $\xi {\cal R}_{f}$, where ${\cal R}_{f}$ 
is
the scalar curvature corresponding to the metric $f_{\alpha\beta}(q)$
and $\xi$ is an arbitrary real constant depending on the operator ordering
used. The resulting
Wheeler-DeWitt equation
reads\cite{GK1}
\begin{equation}
\frac{1}{\rho^{D-1}}\frac{\partial}{\partial\rho}\left(\rho^{D-1}
\frac{\partial\Psi}{\partial\rho}\right)-\frac{1}{\rho^{2}}\sum_{i=1}^{D-1}
\frac{\partial^{2}\Psi}{\partial
z^{i2}}+\left[\xi\frac{(D-1)(D-2)}{\rho^{2}}
-4\omega^{2}\rho^{2}
+4\mu\right]\Psi =0,
\label{WDW}
\end{equation}
where we have used that the scalar curvature of the minisuperspace metric
$f_{\alpha\beta}$ is
${\cal R}_{f}=(D-1)(D-2)\rho^{-2}$.

The inner product for sufficiently regular wavefunctions $\Psi$ and $\Phi$
is defined in the geometric form
\begin{equation}
(\Psi ,\Phi) =\int
d^{D}q\sqrt{|det(f_{\alpha\beta})|}\Psi^{\ast}(q)\Phi (q).
\label{inner}
\end{equation}  

To solve Eq.(\ref{WDW}), let us note that the Hamiltonian commutes with
the generators $-i\frac{\partial}{\partial z^{i}}$ of the symmetry
$z^{i}\rightarrow z^{i}+b^{i}$ with $D-1$ arbitrary real numbers $b^{i}$.
It is
therefore possible to take the solutions of (\ref{WDW}) as eigenstates of
the generators  $-i\frac{\partial}{\partial z^{i}}$
\begin{equation}
\Psi (\rho
,z^{i})=\frac{1}{(2\pi)^{(D-1)/2}}R(\rho)\exp
\left(i\sum_{i=1}^{D-1}F_{i}z^{i}\right),  
\label{Psi-1} 
\end{equation}
where $F_{i}$ is an eigenvalue of the operator $-i\frac{\partial}{\partial
z^{i}}$, which is the quantum version of the conserved quantity defined by
Eq.(\ref{Fi}). The function $R(\rho)$ is then determined as the solution
of the equation 
\begin{equation}
\frac{d^{2}R}{d\rho^{2}}+\frac{D-1}{\rho}\frac{dR}{d\rho}+\frac{K}{\rho^{2}}R
-4\omega^{2}\rho^{2}R=
-4\mu R,
\label{rho-eq}
\end{equation}  
where
\begin{equation}
K\equiv F^{2}+\xi (D-1)(D-2)
\label{K}
\end{equation}
and $F^{2}$ is defined in Eq.(\ref{Ueff}). It is possible to regard
Eq.(\ref{rho-eq}) as a stationary Schrodinger equation in a
$D$-dimensional space (with mass of particle $=1/2$) with an effective
potential
\begin{equation}
U^{(quant)}_{eff}(\rho)\equiv -\frac{K}{\rho^{2}}+4\omega^{2}\rho^{2}.
\label{Uquanteff}
\end{equation}

Depending on whether (i) $K>0$, (ii) $K=0$ or (iii) $K<0$ the effective
potential has an attractive core, is exactly a harmonic potential or has a
repulsive core correspondingly. Then $4\mu$ in Eq.(\ref{rho-eq}) plays the
role of the "energy" eigenvalue. Since the region $\rho\rightarrow\infty$
is classically forbidden, a physically acceptable solution of
Eq(\ref{rho-eq}) has to vanish in this limit. This leads to the
quantization of the eigenvalues of the linear operator corresponding to
Eq.(\ref{rho-eq}). However, in our situation, it is not very appealing to
quantize the energy of dust $\mu$ because this is not a dynamical
variable. In contrast to this, the conserved quantity $F^{2}$, which is a
measure of the anisotropy of the cosmological evolution, can be regarded
as the quantized dynamical variable. Using the fact that $F^{2}$ appears
linearly in Eq.(\ref{rho-eq}), for a {\it given} $\mu$, we select the
appropriate values of $F^{2}$ such that the operator corresponding to the
l.h.s. of Eq.(\ref{rho-eq}) has as its eigenvalue the number $4\mu$.
Therefore we have different values  of $F^{2}$ specifying uniquely the
eigenfunctions. 

The solution of Eq.(\ref{rho-eq}) is given by
\begin{equation}
R=R_{n}(\rho)=N_{n}\rho^{2s_{n}}e^{-|\omega|\rho^{2}}\Phi\left(-n,
\frac{\mu}{|\omega |}-2n, 2|\omega |\rho^{2}\right),
\label{Rn}
\end{equation} 
where  $N_{n}$ is a normalization factor, $\Phi$ is a confluent
hypergeometric function, $n$ is a non-negative
integer and 
\begin{equation}
s_{n}=\frac{1}{4}[-(D-2)+\sqrt{(D-2)^2-4K_{n}}\,]=
\frac{1}{2}\left(\frac{\mu}{|\omega |}-2n-\frac{D}{2}\right).
\label{sn}
\end{equation} 
 The
corresponding quantized values of the length of the vector $F_{i}$ are
\begin{equation}
F^{2}=F^{2}_{n}=4\left\{(D-2)\left[\frac{1}{4}(D-2)-\xi (D-1)\right]-
\left[\frac{\mu}{|\omega |}-1-2n\right]^{2}\right\}.
\label{Fn}
\end{equation}
and $K_{n}$ in (\ref{sn}) is determined by $ F^{2}_{n}$ via 
Eq.(\ref{K}).
Notice that the direction of $F_{i}$ remains arbitrary.

Avoidance of the cosmological singularity in the context of quantum 
cosmology can be defined as a statement that the amplitude 
$\Psi\rightarrow 0$ as $\rho\propto\sqrt{V}\rightarrow 0$. In our case,
 this is possible if $ s_{n}>0$, i.e.
\begin{equation}
\mu>(2n+\frac{D}{2})|\omega |.
\label{avoidance}
\end{equation}
We see that the presence of enough amount of dust is a necessarily
condition for the 
avoidance of the cosmological singularity. It follows from Eq.(\ref{sn})
that condition $s_{n}>0$  implies also that
$K_{n}<0$, that is $F^{2}_{n}<-\xi(D-1)(D-2)$. Assuming $D>2$, this can be
achieved for some $F^{2}_{n}>0$ only if $\xi <0$. Notice that $K_{n}<0$
means that the quantum effective potential (\ref{Uquanteff}) has a repulsive core.

\section{The Dynamics of Quantum Cosmology Proceeding in Cosmic Time }
\subsection{Heisenberg picture}
Before studying the problem of defining the role of time in quantum
cosmology processes in the framework of our model, it is important to
understand some aspects of the classical dynamics. This is of importance
since the quantum behavior has to satisfy, in some way, the correspondence
principle. In particular, the time dependence of $z^{i}$ is determined by
Eq.(\ref{Fi}) which implies
\begin{equation}
\frac{dz^{i}}{dt}=-\frac{F_{i}}{2\rho^{2}}, \quad
i=1,2,...,D-1.
\label{dzidt}
\end{equation}
Eq.(\ref{dzidt}) can be obtained not only from the Euler-Lagrange
equations but also from the Hamiltonian formalism, i.e.
\begin{equation}
\frac{dz^{i}}{dt}=[z^{i},H]_{PB}, \quad \frac{d\pi_{z^{i}}}{dt}=-
\frac{d}{dt}\left(2\rho^{2}\frac{dz^{i}}{dt}\right)=[\pi_{z^{i}},H]_{PB}=0
\label{HamiltForm} 
\end{equation}
where $\pi_{z^{i}}$ is the momenta canonically conjugated to $z^{i}$. Note
that {\it although} $H=0$ {\it along the classical trajectories}, $H$
{\it has nontrivial
Poisson brackets with} $z^{i}$. This allows a nontrivial 
(cosmic)
time dependence through Eqs.(\ref{HamiltForm}), which provides
{\it information
not contained in the constraint equation (\ref{H-diag})}.

We will now see that a similar situation appears in the quantum version of
the theory, where so far the only thing we have cared about has been the 
Wheeler-DeWitt (quantum constraint) equation (\ref{WDW}). Normally, when
quantizing a theory without constraints, the Poisson bracket $[A,B]_{PB}$
is replaced by $i[A,B]$, where $[A,B]$ is the quantum mechanical
commutator of $A$ and $B$. For any operator $Q$ which does not depend
explicitly on time, the Heisenberg equations hold
\begin{equation}
\frac{dQ}{dt}=i[H,Q].
\label{Heis}
\end{equation}

For our purpose, it is convenient to work in the Heisenberg picture, where
operators satisfy Eq.(\ref{Heis}), but states are taken to be time
independent, that is
\begin{equation}
\frac{\partial\Psi}{\partial t}=0.
\label{PsiHeis}
\end{equation}

We are interested in a subspace of this space of functions, the so-called
"physical subspace", in which the constraint equation (\ref{WDW})
holds\cite{Dirac}. The consistency requirement that the constraint be
preserved in time, is here trivially satisfied because the constraint
function $H$ coincides with the Hamiltonian. Solution of Eq.(\ref{WDW})
has been presented in the
previous section, and it is now our purpose to look at the consequences of
the Heisenberg equations that determine {\it the evolution of operators
defined in the whole space of functions} (and not just states satisfying
the constraint $H\Psi =0$).   

The operator equations that govern the cosmic time dependence of $z^{i}$
($i=1,2,...,D-1$) are
\begin{equation}
\frac{dz^{i}}{dt}=i[H,z^{i}]=\frac{i}{4\rho^{2}}\sum_{k=1}^{D-1}
\left[\frac{\partial^{2}}{\partial z^{k2}},z^{i}\right]=
\frac{i}{2\rho^{2}}\frac{\partial}{\partial z^{i}}.
\label{dzidtquant}
\end{equation}

\subsection{Quantum Mechanical Averaging}
We are going now to evaluate the averages of $\frac{dz^{i}}{dt}$ and
other dynamical variables. However, one should be careful on how averages
are defined for wavefunctions (\ref{Psi-1}) since these are normalizable
in a continuous way only.

To see that naive manipulations in this case can lead to wrong results,
let us consider Eq.(\ref{Heis}) together with the subsidiary condition
$H\Psi =0$. If we assume $H$ to be an Hermitian operator, we would get
(using the inner product (\ref{inner})) that 
$\frac{d}{dt}\langle\Psi |Q|\Psi\rangle =\langle\Psi
|\frac{dQ}{dt}|\Psi\rangle =\langle\Psi |i[H,Q]|\Psi\rangle =0$, from
where it
seems to follow that all averages are time independent. However, the above
manipulations are not correct, since the assumption that $H$ is an
Hermitian operator holds only if the functions considered are not
pathological. In fact, the latter is not always true since if we consider,
for
example, the case $Q=z^{i}$, the state $z^{i}|\Psi\rangle $ (that enters
in the expression $\langle\Psi | i[H,Q]|\Psi\rangle$), where
$|\Psi\rangle $ is given by (\ref{Psi-1}), is very singular in the limit
$|z^{i}|\rightarrow\infty$. So, the conclusion about the time independence
of the expectation values of {\it all} canonical operators $Q(q,\pi)$,
i.e. that $\frac{d}{dt}\langle\Psi |Q|\Psi\rangle=0$ is always true, has
been based on a
wrong assumption.

Notice that the $z^{i}$-dependence of the wavefunction  (\ref{Psi-1})
(resulting from the symmetry $z^{i}\rightarrow z^{i}+b^{i}$)
resembles other more familiar example in physics, namely  the momentum 
eigenstate
$|\vec{p}\rangle$ of a
free nonrelativistic particle governed by the Hamiltonian 
$H=\frac{\hat{\vec{p}}^{2}}{2m}$. We have in the Heisenberg picture
$\frac{d}{dt}\langle\vec{p}|\vec{x}|\vec{p}\rangle
=\langle\vec{p}|\frac{d}{dt}\vec{x}|\vec{p}\rangle =
i\langle\vec{p}|[H,\vec{x}]|\vec{p}\rangle =
i\langle\vec{p}|(H\vec{x}-\vec{x}H)|\vec{p}\rangle $. Taking into account
that $H|\vec{p}\rangle =\frac{\vec{p}^{2}}{2m}$ and {\em assuming} $H$ is
hermition, we get $\frac{d}{dt}\langle\vec{p}|\vec{x}|\vec{p}\rangle ?=?
i(\frac{\vec{p}^{2}}{2m}-
\frac{\vec{p}^{2}}{2m})\langle\vec{p}|\vec{x}|\vec{p}\rangle =0$. The
dubious step is actually based on a wrong assumption, when we have taken
for granted the hermiticity of $H$. The condition for the hermiticity of
$H$ fails in this case because of the singular behavior of the state 
$\vec{x}|\vec{p}\rangle$ at large distance and this is the reason why the
result is wrong. In fact, we know that the right answer is obtained by
replacing $i[H,\vec{x}]$ by $\hat{\vec{p}}/m$, that leads to the
conclusion $(\langle\vec{p}|\vec{p}\rangle )^{-1}
\frac{d}{dt}\langle\vec{p}|\vec{x}|\vec{p}\rangle =\vec{p}/m$.
Although the last argument leads to the right result, the method of 
calculation is not totally satisfactory, since it involves carelessly 
manipulating infinite factors like $\langle\vec{p}|\vec{p}\rangle $, etc.
 The same conclusion can however be obtained more rigorously by defining
\begin{equation}
\frac{1}{\langle\vec{p}|\vec{p}\rangle }
\frac{d}{dt}\langle\vec{p}|\vec{x}|\vec{p}\rangle \equiv\lim_{|\varphi\rangle 
  \to |\vec{p}\rangle }\lim_{|\chi\rangle   \to |\vec{p}\rangle }\left[
\frac{1}{\langle\varphi|\chi\rangle }
\frac{d}{dt}\langle\varphi|\vec{x}|\chi\rangle \right]
\label{nonrelat}
\end{equation}
 provided $|\varphi\rangle $ and $|\chi\rangle $ are states in the Heisenberg 
representation which have good enough large distance behavior (e.g. wave 
packets) 
so that the integrals appearing in the numerator and denominator in the
r.h.s. of (\ref{nonrelat})  are well defined (it means that the states 
$|\varphi\rangle $ and $|\chi\rangle $ are not eigenstates of the operators
$\hat{\vec{p}}$ and $H$).

Proceeding in the same fashion in our quantum cosmology model, we define in general
 \begin{equation}
\langle Q\rangle
\equiv\lim_{|\varphi\rangle   \to |\Psi\rangle }
\lim_{|\chi\rangle   \to |\Psi\rangle }\left[
\frac{1}{\langle\varphi|\chi\rangle }
\langle\varphi|Q|\chi\rangle \right], 
\label{correctaverage}
\end{equation}
where $|\Psi\rangle =|F_{i},n\rangle $ is the eigenstate of the operator 
$-i\frac{\partial}{\partial z^{i}}$ with the eigenvalue $ F_{i}$ and the 
corresponding wavefunction $\langle\rho ,z^{i}|\Psi\rangle =
\Psi (\rho ,z^{i})$ is the solution (\ref{Psi-1} ), (\ref{Rn}) of the 
constraint $H\Psi =0$; \quad $|\varphi\rangle $ and $|\chi\rangle $ are 
states in the 
Heisenberg representation (and therefore static) which have good enough 
behavior as $|z^{i}|\rightarrow\infty$, so that the integrals appearing 
in the numerator and denominator of (\ref{correctaverage}) are well defined. 
This necessarily implies that these states do not satisfy the constraint, i.e.
 $H\varphi \neq 0$, $H\chi \neq 0$ although for a given point $(\rho
,z^{i})$, the wavefunctions
$\langle\rho ,z^{i}|\varphi\rangle$ and $\langle\rho ,z^{i}|\chi\rangle$ 
approach the wavefunction $\Psi (\rho ,z^{i})$ in the limit process appearing
in (\ref{correctaverage}). 

In particular, when $Q=\frac{dz^{i}}{dt}$, we find
\begin{equation}
\langle\frac{dz^{i}}{dt}\rangle   =\frac{d\langle z^{i}\rangle}{dt}\equiv
\lim_{|\varphi\rangle   \to |\Psi\rangle }\lim_{|\chi\rangle   \to |\Psi\rangle}
\left[\frac{1}{\langle\varphi|\chi\rangle }
\langle\varphi|\frac{dz^{i}}{dt}|\chi\rangle \right], \,  where \,
|\Psi\rangle =|F_{i},n\rangle .
\label{aver-z-dot}
\end{equation}

Using the definition (\ref{aver-z-dot}) to evaluate the average of (\ref{dzidtquant})
 in a state of the form given by Eq.(\ref{Psi-1}),
we get the result
\begin{equation}
\frac{d\langle z^{i}\rangle }{dt}
=-\langle \frac{1}{2\rho^{2}} \rangle _{n} F_{i},   
\label{aver-z-dot-via-rho}
\end{equation}
which corresponds exactly with the classical result (\ref{dzidt}).

In the same way one can proceed to evaluate other averages. In particular, this 
way of proceeding gives for functions of $\rho$ alone, $Q(\rho)$:
\begin{equation}
\langle Q(\rho) \rangle _{n}=
\frac{\int Q(\rho)R^{2}_{n}(\rho)\rho^{D-1}d\rho}
{\int R^{2}_{n}(\rho)\rho^{D-1}d\rho },  
\label{aver-Q-of-rho}
\end{equation}
where $R_{n}(\rho)$ is determined by Eq.(\ref{Rn}). The convergence of 
$\frac{d\langle z^{i}\rangle }{dt}$ implies the convergence of 
$\langle \frac{1}{\rho^{2}} \rangle $.
One can show
that $\langle \frac{1}{\rho^{2}} \rangle _{n}$ is finite provided 
$\mu >(2n+1)|\omega |$ and
\begin{equation}
\langle \frac{1}{\rho^{2}} \rangle _{n} =
\frac{2\omega^{2}}{\mu -(2n+1)|\omega |}.
\label{aver-1-rho-2}
\end{equation}

 It is interesting to look at the average of $\frac{d^{2}z^{i}}{dt^{2}}$ 
which is given 
(using Heisenberg equations) by
\begin{equation}
\frac{d^{2}z^{i}}{dt^{2}}=-\left(\frac{D-4}{\rho^{4}}+
\frac{2}{\rho^{3}}\frac{\partial}{\partial\rho}\right)
\frac{\partial}{\partial z^{i}}.
\label{d-2z-dt-2}
\end{equation}
It turns out that the average of $\frac{d^{2}z^{i}}{dt^{2}}$ in a state 
given by Eqs.(\ref{Psi-1}) and (\ref{Rn}) is identically zero provided
\begin{equation}
\mu >2(n+1)|\omega |
\label{cond-cons}
\end{equation}
One can show that the restriction (\ref{cond-cons}) on the amount of dust 
is a necessary condition
to 
provide the consistency of 
the main 
three items of our above analyzes:
the constraint equation $H\Psi =0$,  the Heisenberg equations 
and  definition of averages. 

Recall that the
condition (\ref{avoidance}) is needed if we require that the
universe have a zero probability amplitude of having zero volume.  
Comparing (\ref{cond-cons}) with the
condition (\ref{avoidance}), 
we conclude that for $D\geq 4$, the quantum cosmology problem under
consideration has a satisfactory solution if the
condition (\ref{avoidance}) is satisfied. We will assume it in what follows.
 
It is very important also that the
condition (\ref{avoidance}) is a stronger restriction on the amount of
dust
than is actually needed to provide that {\it the average of the "volume"
of the
universe} $\langle V\rangle _{n}=\frac{D}{4(D-1)}\langle \rho^{2}\rangle
_{n}$ is finite and it turns out to be {\it time independent}:
\begin{equation}
\langle V\rangle _{n}=\frac{D}{4(D-1)}
\frac{\int R^{2}_{n}(\rho)\rho^{D+1}d\rho}
{\int R^{2}_{n}(\rho)\rho^{D-1}d\rho }=
\frac{D}{16(D-1)\omega^{2}}\frac{\mu^{2}-(2n+1)^{2}\omega^{2}}
{\mu -(2n-1)|\omega|} 
\label{V-average}
\end{equation}
In particular, for the ground state, $n=0$, $\langle V\rangle _{n=0}=
\frac{D}{16(D-1)\omega^{2}}(\mu -|\omega|)$. The integrals appearing in 
(\ref{V-average}) when they converge of course give a positive result
as it must be from the definition of $\langle V\rangle _{n}$.

\subsection{The Cosmic Time Dependence of the Expectation Values of the
Cosmological Quantities: Some General Results. }

To see more clear the above results, let us now represent
them in terms of usual cosmological quantities. In the totally anisotropic
 model (\ref{metric}), the corresponding classical variables are the 
"scale factors" $a_{l}(t)$ \, ($l=1,2,...,D$) which have been parametrized
 by means of Eq.(\ref{teta-param}) in terms of the "volume" of the universe 
$V(t)$ and
$D$ functions $\theta_{l}(t)$ \, ($l=1,2,...,D$). Remind that due to the identity
$\sum_{l=1}^{D}\theta_{l}\equiv 0$, only $D-1$ of the $\theta_{l} $'s are 
independent . Relations between variables $V$ and $\theta_{i} $ 
($i=1,2,...,D-1$) and those 
$\rho$ and $z^{i}$ are given by Eqs.(\ref{rho}) and (\ref{z}).

In quantum cosmology, the average of $z^{i}$, even with the improved definition
(\ref{correctaverage}), does not exist and therefore (\ref{correctaverage}) has
 nothing to say on whether $\langle z_{i} \rangle$ is time dependent or
time
 independent. In contrast to this, the average of 
$\langle\frac{dz^{i}}{dt}\rangle $ is well defined and in the present of an 
anisotropy ($F_{i}\neq 0$), it is a nonzero finite constant determined by
Eqs.(\ref{aver-z-dot-via-rho}) and (\ref{aver-1-rho-2}). Then we find that
the time evolution of  $\langle z_{i} \rangle $ is of the form 
$\langle z_{i} \rangle =\langle\frac{dz^{i}}{dt}\rangle t+c^{i}= 
\left(-\langle \frac{1}{2\rho^{2}} \rangle _{n} F_{i}\right)t+c^{i}$, where
$c^{i}$ are undetermined constants. This yields to the following (cosmic)
time dependence
of the $\theta_{i}$ variables
\begin{equation}
\langle \theta_{i} \rangle =\alpha_{i}t +\gamma_{i}, 
\label{theta-t}
\end{equation}
where $\gamma_{i}$ are integration constants and
for $i=1,2,...,D-1$
\begin{equation}
\alpha_{i}=\frac{\sqrt{D-1}}{D(\sqrt{D}+1)}
\frac{2\omega^{2}}{\mu -(2n+1)|\omega |}
\left[(D+\sqrt{D}-1)F_{i}-\sum_{j\neq i}F_{j}\right]
\label{alpha-i}
\end{equation}
From the identity $\sum_{l=1}^{D}\theta_{l}\equiv 0$ we have
for $\langle \theta_{D} \rangle $
\begin{equation}
\langle \theta_{D} \rangle =\alpha_{D}t +\gamma_{D} \quad where 
\quad
\alpha_{D}=-\sum_{i=1}^{D-1}\alpha_{i}, \quad 
\gamma_{D}=-\sum_{i=1}^{D-1}\gamma_{i}
\label{theta-D}
\end{equation}

Besides, in the classical cosmology one can define  $D$ expansion
 parameters
$H_{l}\equiv\frac{1}{a_{l}(t)} \frac{d a_{l}(t)}{dt}$. 
In the quantum version of the theory, we have of course to define 
the ordering of the operators $a_{l}$ and $\frac{d a_{l}}{dt}$. For
 example, $a_{l}^{-1}\frac{d a_{l}}{dt}$ or 
$a_{l}^{-1/2}\frac{d a_{l}}{dt}a_{l}^{-1/2}$, etc. give 
quantum mechanically distinct definitions for $H_{l}$. We will choose
a definition
\begin{equation}
H_{l}\equiv\frac{d}{dt}(\ln a_{l}).
\label{h-l-defin}
\end{equation}
Using this definition, the parametrization 
(\ref{teta-param}) and results of the previous subsection one can 
evaluate {\it the expectation values of the expansion parameters}
$\langle H_{l} \rangle $ {\it which turn out to be constants}:
\begin{eqnarray}
\langle H_{l} \rangle &=&\left(\frac{D}{4(D-1)}\right)^{1/D}
\left(\langle \frac{d}{dt}(\ln\rho^{2/D}) \rangle+
\langle \frac{d\theta_{l}}{dt} \rangle \right)
\nonumber \\
 &=& \left(\frac{D}{4(D-1)}\right)^{1/D}
\langle \frac{d\theta_{l}}{dt} \rangle 
=\left(\frac{D}{4(D-1)}\right)^{1/D}\alpha_{l},
\label{h-l-aver}
\end{eqnarray}
where $\alpha_{l}$ are determined in (\ref{alpha-i}) and
 (\ref{theta-D}). 

Let us notice that Eqs.(\ref{teta-param}) and (\ref{h-l-defin})
together with the identity $\sum_{l=1}^{D}\theta_{l}\equiv 0$
imply that
\begin{equation}
\sum_{l=1}^{D}H_{l}=\frac{d}{dt}(\ln V).
\label{sum h-l}
\end{equation}

Taking average on both sides of Eq.(\ref{sum h-l}), we obtain the result 
that {\it the sum of the averages of the expansion parameters equals
zero}, due to Eqs.(\ref{h-l-aver}) and (\ref{theta-D}) or alternatively 
from Eqs.(\ref{aver-Q-of-rho}) and (\ref{V-average}). This shows that the 
definition (\ref{h-l-defin}) of the expansion parameters is consistent
with the quantum stabilization of the volume of the universe, 
$\langle V \rangle =const$,  
Eq.(\ref{V-average}).

From (\ref{h-l-defin}) and (\ref{h-l-aver}), we see that the time 
behavior of $\langle \ln a_{l} \rangle $ is given by
\begin{equation}
\langle \ln a_{l} \rangle
 =\left(\frac{D}{4(D-1)}\right)^{1/D}
\alpha_{l}t+\tilde{\gamma}_{l}, \quad l=1,2,...,D,
\label{ln-a-aver}
\end{equation}
where $\tilde{\gamma}_{l}$ are arbitrary integration constants.

\section{Inflation-Compactification as a Quantum Effect}

We will now see that
the results of Sec.4 allow to realize a dynamical explanation of the
asymmetry in the sizes of extra and ordinary dimensions in the context of
quantum cosmology.

At the classical level, there is no difference on whether we use
$a_{l}$ or $\ln a_{l}$ as our variables. If Eq.(\ref{ln-a-aver})
were to hold classically, we could conclude that some dimensions   
exponentially expand and others exponentially contract, depending on the
sign of $\alpha_{l}$, as given by (\ref{alpha-i}) and
 (\ref{theta-D}). In our case, the behavior of the universe is 
intrinsically quantum mechanical and we will refer to a "{\it quantum
inflationary phase}" for a given dimension $l$ if the expectation value of
the expansion parameter $\langle H_{l} \rangle =const >0$. Likewise we
will refer to a "{\it quantum deflationary phase}" for a given dimension
$l$ if
the expectation value of
the expansion parameter $\langle H_{l} \rangle =const <0$.

A case of particular interest is when the expectation values of the
expansion parameters of three of the dimensions are identical and   
at the same time the  expectation values of the
expansion parameters of the remaining $D-3$ dimensions are also
identical. In such a case $\alpha_{1}=\alpha_{2}=\alpha_{3}\equiv
\alpha$ and $\alpha_{4}=\alpha_{5}=...=\alpha_{D}\equiv
\tilde{\alpha}$. Then it follows from the identity
$\sum_{l=1}^{D}\alpha_{l}=0$ that $\tilde{\alpha}=-3\alpha/(D-3)$  
and we get by using Eq.(\ref{alpha-i}):
\begin{equation}
\alpha =\alpha(n)=\pm\frac{2\omega^{2}}{D[\mu -(2n+1)|\omega|]}
\sqrt{\frac{(D-3)(D-1)}{3}}|F|_{n},
\label{alpha-n-infl}
\end{equation}
\begin{equation}
\tilde{\alpha} =\tilde{\alpha}(n)=\mp\frac{2\omega^{2}}{D[\mu
-(2n+1)|\omega|]}
\sqrt{\frac{3(D-1)}{D-3}}|F|_{n},
\label{alpha-n-comp}
\end{equation}
where $|F|_{n}\equiv\sqrt{F^{2}_{n}}$ and $ F^{2}_{n}$ is determined
by Eq.(\ref{Fn}).

Invoking our definitions of quantum inflationary phase and of
quantum deflationary phase, we observe that one set of dimensions is in a
quantum inflationary phase and simultaneously another set of dimensions
is in quantum deflationary phase. This situation is described by the
following equations:
\begin{eqnarray}
\langle H_{i} \rangle &=&\alpha(n) \quad for \quad i=1,2,3;
\nonumber
\\
\langle H_{j} \rangle &=&\tilde{\alpha}(n) \quad for \quad j=4,...,D.
\label{infl-comp}
\end{eqnarray}
According to Eqs.(\ref{alpha-n-infl}) and (\ref{alpha-n-comp}), choosing
the three dimensional subspace to be expanding one, we get the
simultaneous
contraction of the extra dimensions. During this
quantum "inflation-compactification" process, the expectation value of the
"volume" of the universe, $V$, remains constant determined by
Eq.(\ref{V-average}).

\section{Discussion}
The minisuperspace model of quantum cosmology we discussed here
demonstrates a very interesting
feature which is absent, as far as we know, in all other known quantum
cosmology models. Namely, a widespread belief that the
cosmic time, which one uses in classical cosmology, disappears in quantum
cosmology altogether, seems to be not always right. In the presented model
we have seen that  quantum mechanical averages of certain cosmological
quantities can explicitly depend on the same cosmic time which was used in
the
appropriate classical cosmological model. Short explanation of the essence
of the idea was given in the introduction, and for technical details see
Sec.4
and Ref.$[8]$. Notice that the anisotropy in the  evolution of the
universe is an essential element which provides this unique feature of
 our Kaluza-Klein model.

It has been found that quantum effects stabilize the volume of the
universe, so that there can be avoidance of the cosmological singularity.
The stabilization of volume is consistent with a new quantum effect:
existence of a quantum inflationary phase for some dimensions  and
simultaneous quantum deflationary phase for the remaining dimensions.
This effect can be responsible for a visible asymmetry between ordinary
and extra dimensions.
One can show\cite{GK2} that the above results also follow if instead of
dust we introduce a massive scalar field whose homogeneous degrees of
freedom are described quantum mechanically.

\end{document}